\documentclass{article}

    \usepackage[preprint, nonatbib]{neuripsDeepInverse_2021}

\usepackage[utf8]{inputenc} 
\usepackage[T1]{fontenc}    
\usepackage{hyperref}       
\usepackage{url}            
\usepackage{booktabs}       
\usepackage{amsfonts}       
\usepackage{nicefrac}       
\usepackage{microtype}      
\usepackage{amsthm}
\usepackage{amssymb}
\usepackage{amsmath}
\usepackage{amsfonts}
\usepackage{mathtools}

\usepackage[sorting=nyt]{biblatex}
\def \x {\bf x}

\def \R {\mathbb{R}}
\def \xhat {\mathbf{\hat{x}}}

\newtheorem{theorem}{Theorem}[section]
\newtheorem{corollary}{Corollary}[theorem]
\newtheorem{lemma}[theorem]{Lemma}
\newtheorem{proposition}[theorem]{Proposition}

\title{Beyond Independent Measurements: General Compressed Sensing with GNN Application}

\author{
  Alireza Naderi \\
  Department of Mathematics\\
  University of British Columbia\\
  Vancouver, BC V6T 1Z2 \\
  \texttt{alireza@math.ubc.ca} \\
   \And
   Yaniv Plan \\
   Department of Mathematics \\
   University of British Columbia \\
   Vancouver, BC V6T 1Z2 \\
   \texttt{yaniv@math.ubc.ca} \\
}

\begin{document}

\maketitle

\begin{abstract}
We consider the problem of recovering a structured signal $\mathbf{x} \in \mathbb{R}^{n}$ from noisy linear observations $\mathbf{y} =\mathbf{M} \mathbf{x}+\mathbf{w}$. The measurement matrix is modeled as $\mathbf{M} = \mathbf{B}\mathbf{A}$, where $\mathbf{B} \in \mathbb{R}^{l \times m}$ is arbitrary and $\mathbf{A} \in \mathbb{R}^{m \times n}$ has independent sub-gaussian rows. By varying $\mathbf{B}$, and the sub-gaussian distribution of $\mathbf{A}$, this gives a family of measurement matrices which may have heavy tails, dependent rows and columns, and singular values with a large dynamic range. When the structure is given as a possibly non-convex cone $T \subset \mathbb{R}^{n}$, an approximate empirical risk minimizer is proven to be a robust estimator if the effective number of measurements is sufficient, even in the presence of a model mismatch. In classical compressed sensing with independent (sub-)gaussian measurements, one asks \textit{how many measurements are needed to recover $\mathbf{x}$?} In our setting, however, the effective number of measurements depends on the properties of $\mathbf{B}$. We show that the \textit{effective rank} of $\mathbf{B}$ may be used as a surrogate for the number of measurements, and if this exceeds the squared \textit{Gaussian mean width} of $(T-T) \cap \mathbb{S}^{n-1}$, then accurate recovery is guaranteed. Furthermore, we examine the special case of generative priors in detail, that is when $\mathbf{x}$ lies close to $T = \mathrm{ran}(G)$ and $G: \mathbb{R}^k \rightarrow \mathbb{R}^n$ is a Generative Neural Network (GNN) with ReLU activation functions.  Our work relies on a recent result in random matrix theory by Jeong, Li, Plan, and Y{\i}lmaz \cite{jeong2020subgaussian}.
\end{abstract}

\section{Introduction} \label{intro}
\par

\par
In compressed sensing \cite{foucart2013mathematical}, the goal is to reconstruct a high-dimensional signal $\mathbf{x} \in \mathbb{R}^n$ from a noisy low-dimensional linear transformation of it, $\mathbf{y} = \mathbf{M}\mathbf{x} + \mathbf{w} \in \mathbb{R}^l,\, l < n$. Even in the absence of noise, the reconstruction would not be possible without a further assumption of \textit{signal structure}, i.e., some restriction of the possible values of $\mathbf{x}$.  In early works in compressed sensing, the signal structure was encoded via sparsity, or sparsity with respect to a dictionary. A sufficient condition on the \textit{measurement matrix} $\mathbf{M}$ that enables robust recovery is the celebrated Restricted Isometry Property (RIP) \cite{candes2008restricted}. Intriguingly, all known algorithms for certifying RIP either are computationally intractable or only function in the parameter regimes which are highly suboptimal \cite{ding2020average}, but a sub-gaussian measurement matrix with independent rows satisfies the RIP with high probability in near optimal parameter regimes \cite{baraniuk2008simple}.

More recently, compressed sensing ideas have been generalized to allow the signal structure to be almost arbitrary, encoded as a subset $T \subset \R^n$, provided that the measurements are appropriately random.  Gaussian measurement matrices are rotationally invariant \cite{vershynin}, which allows them to be universally effective for general compressed sensing.  This was proven using \cite{chandrasekaran2012convex, rudelson2008sparse} or generalizing \cite{stojnic2013framework, oymak2013squared, thrampoulidis2014simple} Gordon's theorem \cite{gordon1988milman}, and also through a foundational result in conic geometry \cite{amelunxen2014living}. Sub-gaussian matrices, as defined later, are essentially the largest class of matrices that approximately satisfy rotation invariance, and also enjoy universal guarantees, although with different proof techniques relying on chaining arguments \cite{mendelson2007reconstruction, liaw2017simple, jeong2020subgaussian}.  We also note a few results outside of the Gaussian or sub-gaussian framework: \cite{tropp2015convex,oymak2018universality, mendelson2014learning, mendelson2010empirical}.  

The recent results of \cite{jeong2020subgaussian} extend previous theory by allowing the measurement matrix to take the form $\mathbf{M} = \mathbf{B}\mathbf{A}$ with arbitrary $\mathbf{B}$ and sub-gaussian $\mathbf{A}$ with independent rows.  They prove a general restricted isometry property,    
%
%
thereby giving basic signal recovery guarantees for measurement matrices with dependent rows.  This random model is compelling because real-data measurement matrices often have highly variable singular values, but sub-gaussian matrices with independent rows typically have nearly uniform singular values.  In fact, under mild assumptions, the singular values of $\mathbf{B A}$ concentrate around the singular values of $\mathbf{B}$ (after rescaling), and so, one may target any singular value vector by adjusting the singular values of $\mathbf{B}$.
We build upon the work of \cite{jeong2020subgaussian} to show accurate signal recovery by the generalized Lasso, even in the presence of inexact optimization and/or a mismatch in the signal structure. We believe this provides the first general compressed sensing guarantees allowing any two of the following three items simultaneously (we allow all three): (i) the measurement matrix need not have independent rows, (ii) the signal may be estimated by generalized Lasso, but without exact optimization, and (iii) the signal is allowed to be only \textit{approximately} structured, i.e. close to the structure set and not belonging to it.  The improved dependence on the sub-gaussian parameter of the measurement matrix in \cite{jeong2020subgaussian} is reflected in our work as well.

\par
Items (ii) and (iii) above make our framework well-suited to the setting in which the signal structure is the range of a neural network.  Indeed, in recent years, it has been shown empirically that learning the appropriate signal structure by fitting it to the range of a neural network can be much more effective than using a predetermined structure such as sparsity \cite{bora2017compressed, van2018compressed, jalal2020robust}. In this case, one assumes that $T$ is the range of a generative model $G: \R^k \rightarrow \R^n$ which is already trained.  While training often approximates the signal structure well, any new signal outside of the training set will typically be close, but not in $T$; thereby necessitating item (iii).  The standard method of estimating $\x$ is to find the restricted least squares fit, i.e., to let the estimate $\xhat$ be the solution to the program, sometimes called generalized Lasso:
\begin{equation}\tag{P} \label{program}
    \mathrm{minimize} \; ||\mathbf{y} - \mathbf{M}\mathbf{x}'||_2^2 \quad \mathrm{s.t.} \; \mathbf{x}' \in T.
\end{equation}
Since $T = \mathrm{ran}(G)$ is non-convex, and no convex relaxation of such a program is known in general, typical gradient-descent-based algorithms would not be guaranteed to approach a global minimum; thereby necessitating item (ii). Our theory requires an upper bound of the Gaussian mean width of ${(T-T) \cap \mathbb{S}^{n-1}}$, which we give in Proposition \ref{width} below; we believe this is novel.

\section{Main Results}

\subsection{Problem Setup} \label{setup}
\par

\par
Recall that a random variable $Z$ is called \textit{sub-gaussian}, if its tail probability is dominated by that of a Gaussian random variable. An equivalent rigorous definition requires the \textit{sub-gaussian norm} to be finite: $||Z||_{\psi_2} \coloneqq \inf \big\{t > 0 \, \big | \; \mathbb{E}(Z^2/t^2) \leq 2 \big\} < \infty $. A random vector $\mathbf{v} \in \mathbb{R}^n$ is \textit{sub-gaussian}, if all of its one-dimensional marginals are sub-gaussian random variables. Mathematically, if $||\langle \mathbf{\theta}, \mathbf{v} \rangle||_{\psi_2}$ is finite for all $\mathbf{\theta} \in \mathbb{S}^{n-1}$, we define $ ||\mathbf{v}||_{\psi_2} \coloneqq \sup_{\mathbf{\theta} \in \mathbb{S}^{n-1}} ||\langle \mathbf{\theta}, \mathbf{v} \rangle||_{\psi_2}$.

We will consider a random matrix $\mathbf{A} \in \mathbb{R}^{m \times n}$ whose rows $\{\mathbf{a}_1^\top, \cdots, \mathbf{a}_m^\top\}$ are statistically independent, mean-zero ($\mathbb{E}\mathbf{a}_i = \mathbf{0}$), isotropic ($\mathbb{E} \mathbf{a}_i \mathbf{a}_i^\top = \mathbf{I}_n$), and sub-gaussian with parameter $K$ ($||\mathbf{a}_i||_{\psi_2} \leq K$). We let the measurement matrix be the product $\mathbf{B}\mathbf{A}$, where $\mathbf{B} \in \mathbb{R}^{l \times m}$ is arbitrary. In other words, we let every row of our measurement matrix be an arbitrary linear combination of $\mathbf{a}_1^\top, \cdots, \mathbf{a}_m^\top$. Our goal is to determine the criteria that $\mathbf{B}$, $\mathbf{A}$, and the structure set $T$ must satisfy for accurate recovery to be possible. Informally, one requires $\mathbf{B}$ to be far from low-rank, otherwise, the number of independent effective measurements would not be sufficient. An appropriate quantity is the \textit{stable rank} defined as
\begin{equation}
    \mathrm{sr}(\mathbf{B}) \coloneqq \frac{||\mathbf{B}||_F^2}{||\mathbf{B}||^2} = \frac{\sum_{i=1}^{\mathrm{rank}(\mathbf{B})} \sigma_i^2}{\max_{i} \sigma_i^2} \leq \mathrm{rank}(\mathbf{B}),
\end{equation}
where $\sigma_i$'s are the singular values of $\mathbf{B}$. The main advantage of this definition over rank itself is that it is robust to the small non-zero singular values that increase the rank but do not contribute to an effective measurement, hence the name stable rank. Furthermore, one requires $T$ to be \textit{small} in some sense, ideally not to intersect the null space of $\mathbf{M} = \mathbf{B}\mathbf{A}$. A useful notion of size is the \textit{Gaussian mean width} defined as
\begin{equation}
    w(T) \coloneqq \mathbb{E} \sup_{\mathbf{v} \in T} \langle \mathbf{v}, \mathbf{g} \rangle,
\end{equation}
where the expectation is calculated with respect to $\mathbf{g} \sim \mathcal{N}(\mathbf{0}, \mathbf{I}_n)$. The reader may consult \cite{plan2012robust} for some basic properties of Gaussian mean width. The following theorem gives the desired estimation bound.

\subsection{Main Theorem} \label{mainthm}
\begin{theorem} \label{main}
Let $\mathbf{x} \in \mathbb{R}^n$, $\mathbf{B} \in \mathbb{R}^{l \times m}$ be an arbitrary fixed matrix, and $\mathbf{A} \in \mathbb{R}^{m \times n}$ be a matrix whose rows are independent, mean-zero, isotropic, and sub-gaussian vectors with sub-gaussian parameter $K$. Let $T \subset \mathbb{R}^n$ be a closed cone and define $T' \coloneqq (T-T)\cap \mathbb{S}^{n-1}$. Let $\mathbf{y} = \mathbf{B}\mathbf{A}\mathbf{x}+\mathbf{w}$ for some fixed unknown $\mathbf{w} \in \mathbb{R}^m$. Let $\xhat \in T$ satisfy $||\mathbf{y} - \mathbf{B}\mathbf{A} \xhat||_2^2 \leq \min_{\mathbf{x}' \in T} ||\mathbf{y} - \mathbf{B}\mathbf{A}\mathbf{x}'||_2^2 + \epsilon^2$. If $\mathrm{sr}(B) \gg K^2 \log K \cdot w^2(T')$, then with probability larger than $1-11e^{-w^2(T')}$,
\begin{equation} \label{maineq}
    ||\mathbf{x} - \xhat||_2 \lesssim \frac{Kw(T')}{||\mathbf{B}||_F \sqrt{\mathrm{sr}(\mathbf{B})}} ||\mathbf{w}||_2 + \frac{\epsilon}{||\mathbf{B}||_F} + \frac{K\sqrt{l}}{\sqrt{\mathrm{sr}(\mathbf{B})}} \mathrm{dist}(\mathbf{x}, T).
\end{equation}
\end{theorem}

\par
There are three sources of error present here: the additive noise $\mathbf{w}$, the inaccuracy in optimization $\epsilon$, and the model mismatch, i.e., $\mathbf{x}$ not exactly belonging to $T$. Note that ${\mathrm{sr}(\mathbf{B}) \leq \mathrm{rank}(\mathbf{B}) \leq \min(l,m)}$, so basically the effective number of measurements is bounded by the number of underlying independent measurements $m$. When $\mathbf{B}$ is a multiple of identity, the equality case $\mathrm{sr}(\mathbf{B}) = m$ occurs.

\par
What happens if we have more effective measurements than needed? The energy of the noise that appears in the total error in (\ref{maineq}) is decreased by the \textit{oversampling factor} ${\mathrm{sr}(\mathbf{B})}/w^2(T')$. In other words, if we have some control over the measurement matrix, then we are able to denoise the original signal by increasing $\mathrm{sr}(\mathbf{B})$. In case of $\mathbf{B} = \mathbf{I}_m$, this simply means increasing the number of measurements $m$. This denoising effect is well known for stochastic noise, but we believe that it had previously been shown only for non-random noise under the assumption of a Gaussian measurement matrix, as in \cite{thrampoulidis2014simple}.  We note, it is important here that the noise is \textit{fixed} i.e., not chosen adversarially depending on the realization of $\mathbf{A}$.

\section{Application on Generative Neural Networks} \label{app}
\par

Due to the success of deep generative neural networks to learn complex structures, recent works on compressed sensing have considered generative priors instead of the traditional sparsity \cite{bora2017compressed}. While some works require the trained network $G: \mathbb{R}^k \rightarrow \mathbb{R}^n$ to be $L$-Lipschitz \cite{bora2017compressed} and prove that $\mathcal{O}(k \log L)$ number of measurements would suffice for a recovery guarantee, a drawback of such analysis is that the Lipschitz constant of the network cannot be calculated solely based on its architecture. Here we present a different approach: we give an upper bound of the Gaussian mean width of $T' = \big(\mathrm{ran}(G) - \mathrm{ran}(G)\big) \cap \mathbb{S}^{n-1}$, which only depends on the hyperparameters of the model. Then we apply Theorem \ref{main} to achieve a recovery guarantee for compressed sensing with generative priors.

\subsection{GNN and Guassian Mean Width} \label{GMW}
\par
    A $d$-layer GNN with ReLU activation function is a function $G: \mathbb{R}^k \rightarrow \mathbb{R}^n$ of the form
    \begin{equation} \label{gnn}
        G(\mathbf{z}) = \sigma(\mathbf{A}_d \sigma(\mathbf{A}_{d-1} \sigma( \dots \mathbf{A}_2 \sigma (\mathbf{A}_1 \mathbf{z}) \dots ))),
    \end{equation}
where $\sigma(\cdot) = \max(\cdot, 0)$ is applied entrywise and $\mathbf{A}_i \in \mathbb{R}^{p_i \times p_{i-1}}$, $p_0 = k$, and $p_d = n$. The weight matrices $\mathbf{A}_i$ are sometimes assumed to have iid Gaussian entries, however, our analysis puts no requirement on $\mathbf{A}_i$. To bound the Gaussian mean width of $\mathrm{ran}(G)$, we first show that the set is contained in a union of subspaces with minimal count. 
Note that the choice of ReLU activation function is not essential to what follows, and any piecewise linear function with only two pieces (e.g. leaky ReLU) could be substituted.  Similar counting arguments have appeared in various theoretical works on expansive neural nets.

\begin{lemma} \label{count}
Let $G$ be as in (\ref{gnn}) and $T = \mathrm{ran}(G)$. We have $T \subset \bigcup_{i=1}^{N} E_i$,
where each $E_i$ is a subspace of dimension at most $k$ and
\begin{equation}
    N \leq \Big[\big(\frac{2e}{k}\big)^{d}\Big(\prod_{i=1}^{d}p_i\Big)\Big]^k.
\end{equation}
Consequently,
\begin{equation*}
    T-T \subset \bigcup_{i=1}^{N \choose 2} F_i,
\end{equation*}
where each $F_i$ is a subspace of dimension at most $2k$.
\end{lemma}

\begin{proposition} \label{width}
Let $G$ be as in (\ref{gnn}) and $T = \mathrm{ran}(G) \subset \mathbb{R}^n$. Then the following holds:
\begin{equation}
    w \big( T \cap \mathbb{S}^{n-1} \big) \leq w \big( (T-T) \cap \mathbb{S}^{n-1} \big) \lesssim \sqrt{kd \log \Big( \frac{p'}{k} \Big)},
\end{equation}
where $p' = \Big( \prod_{j=1}^{d} p_j \Big)^{1/d}$ is the geometric mean of $p_1, \cdots, p_d$.
\end{proposition}
\par

\subsection{Compressed Sensing with Generative Priors} \label{CS-GNN}

The effective number of measurements required for any recovery algorithm to be successful is $\mathcal{O} \big( w^2(T') \big) = \mathcal{O} \big( kd \log (p'/k) \big)$. Typically, a GNN is assumed to be expansive, that is $k = p_0 \leq p_1 \leq \cdots \leq p_d = n$. Thus, $k \leq p' \leq n$, which slightly improves $\mathcal{O}(kd \log n)$ in \cite{bora2017compressed}. Moreover, by combining Theorem \ref{main} and Proposition \ref{width}, we present a recovery guarantee for the compressed sensing with generative priors that not only allows for a much broader range of measurement matrices, but also demonstrates the dependence on their parameters.

\begin{corollary}
Consider the settings of Theorem \ref{main} for $T = \mathrm{ran}(G)$ as in \eqref{gnn}. If $\mathrm{sr}(B) \gg K^2 \log K \cdot kd \log(p'/k)$, then
\begin{equation}
    ||\mathbf{x} - \xhat||_2 \lesssim \frac{K\sqrt{kd \log(p'/k)}}{||\mathbf{B}||_F \sqrt{\mathrm{sr}(\mathbf{B})}} ||\mathbf{w}||_2 + \frac{\epsilon}{||\mathbf{B}||_F} + \frac{K\sqrt{l}}{\sqrt{\mathrm{sr}(\mathbf{B})}} \mathrm{dist}\big(\mathbf{x}, \mathrm{ran}(G)\big).
\end{equation}
\end{corollary}

\par
Our analysis improves upon the best known results in three ways: Firstly, our result suggests a \textit{denoising} behaviour as the number of measurements (or the stable rank of $\mathbf{B}$) increases, even though the noise is not assumed to be random. Intuitively, one expects a smaller error bound associated with less compression, and our theory highlights this dependence. In contrast, the bound in \cite{bora2017compressed} shows constant dependence on the noise level, the optimization margin, and the model mismatch, regardless of the number of measurements. Secondly, we improve the logarithmic factor in the compression bound, i.e. we need $\mathcal{O}(kd \log(p'/k)) \leq \mathcal{O}(kd \log(n/k))$ effective measurements, rather than $\mathcal{O}(kd \log n)$. This was also available to the authors of \cite{bora2017compressed}, had they used a tighter inequality in proof of Lemma 8.3. Finally, we require milder conditions for the measurement matrix, allowing for a fixed \textit{mixing} matrix to create dependence among the rows. This improvement is based on the geometry-preserving properties discussed in \cite{jeong2020subgaussian}.

\section{Summary} \label{conclude}
\par
We achieve an estimation bound for the reconstruction of structured signals using noisy and dependent random measurements. The generality of our model enables its application on the non-traditional structure sets such as the range of a ReLU generative neural network. We believe that this is the first general compressed sensing result that specializes well to the generative structure. Whether a similar recovery guarantee is obtainable for GNNs with other activation functions (e.g. sigmoid or tanh), remains an open question.

\newpage
\printbibliography

@inproceedings{thrampoulidis2014simple,
  title={Simple error bounds for regularized noisy linear inverse problems},
  author={Thrampoulidis, Christos and Oymak, Samet and Hassibi, Babak},
  booktitle={2014 IEEE International Symposium on Information Theory},
  pages={3007--3011},
  year={2014},
  organization={IEEE}
}

@article{mendelson2010empirical,
  title={Empirical processes with a bounded $\psi_1$ diameter},
  author={Mendelson, Shahar},
  journal={Geometric and Functional Analysis},
  volume={20},
  number={4},
  pages={988--1027},
  year={2010},
  publisher={Springer}
}

@article{oymak2018universality,
  title={Universality laws for randomized dimension reduction, with applications},
  author={Oymak, Samet and Tropp, Joel A},
  journal={Information and Inference: A Journal of the IMA},
  volume={7},
  number={3},
  pages={337--446},
  year={2018},
  publisher={Oxford University Press}
}

@inproceedings{mendelson2014learning,
  title={Learning without concentration},
  author={Mendelson, Shahar},
  booktitle={Conference on Learning Theory},
  pages={25--39},
  year={2014},
  organization={PMLR}
}

@incollection{liaw2017simple,
  title={A simple tool for bounding the deviation of random matrices on geometric sets},
  author={Liaw, Christopher and Mehrabian, Abbas and Plan, Yaniv and Vershynin, Roman},
  booktitle={Geometric aspects of functional analysis},
  pages={277--299},
  year={2017},
  publisher={Springer}
}

@article{mendelson2007reconstruction,
  title={Reconstruction and subgaussian operators in asymptotic geometric analysis},
  author={Mendelson, Shahar and Pajor, Alain and Tomczak-Jaegermann, Nicole},
  journal={Geometric and Functional Analysis},
  volume={17},
  number={4},
  pages={1248--1282},
  year={2007},
  publisher={Springer}
}

@article{rudelson2008sparse,
  title={On sparse reconstruction from Fourier and Gaussian measurements},
  author={Rudelson, Mark and Vershynin, Roman},
  journal={Communications on Pure and Applied Mathematics: A Journal Issued by the Courant Institute of Mathematical Sciences},
  volume={61},
  number={8},
  pages={1025--1045},
  year={2008},
  publisher={Wiley Online Library}
}

@inproceedings{oymak2013squared,
  title={The squared-error of generalized lasso: A precise analysis},
  author={Oymak, Samet and Thrampoulidis, Christos and Hassibi, Babak},
  booktitle={2013 51st Annual Allerton Conference on Communication, Control, and Computing (Allerton)},
  pages={1002--1009},
  year={2013},
  organization={IEEE}
}

@article{chandrasekaran2012convex,
  title={The convex geometry of linear inverse problems},
  author={Chandrasekaran, Venkat and Recht, Benjamin and Parrilo, Pablo A and Willsky, Alan S},
  journal={Foundations of Computational mathematics},
  volume={12},
  number={6},
  pages={805--849},
  year={2012},
  publisher={Springer}
}

@article{stojnic2013framework,
  title={A framework to characterize performance of lasso algorithms},
  author={Stojnic, Mihailo},
  journal={arXiv preprint arXiv:1303.7291},
  year={2013}
}

@article{amelunxen2014living,
  title={Living on the edge: Phase transitions in convex programs with random data},
  author={Amelunxen, Dennis and Lotz, Martin and McCoy, Michael B and Tropp, Joel A},
  journal={Information and Inference: A Journal of the IMA},
  volume={3},
  number={3},
  pages={224--294},
  year={2014},
  publisher={OUP}
}

@book{vershynin,
  title={High-dimensional probability: An introduction with applications in data science},
  author={Vershynin, Roman},
  volume={47},
  year={2018},
  publisher={Cambridge university press}
}

@inproceedings{bora2017compressed,
  title={Compressed sensing using generative models},
  author={Bora, Ashish and Jalal, Ajil and Price, Eric and Dimakis, Alexandros G},
  booktitle={International Conference on Machine Learning},
  pages={537--546},
  year={2017},
  organization={PMLR}
}

@article{plan2012robust,
  title={Robust 1-bit compressed sensing and sparse logistic regression: A convex programming approach},
  author={Plan, Yaniv and Vershynin, Roman},
  journal={IEEE Transactions on Information Theory},
  volume={59},
  number={1},
  pages={482--494},
  year={2012},
  publisher={IEEE}
}

@article{jeong2020subgaussian,
  title={Sub-Gaussian matrices on sets: Optimal tail dependence and applications},
  author={Jeong, Halyun and Li, Xiaowei and Plan, Yaniv and Y{\i}lmaz, {\"O}zg{\"u}r},
  journal={arXiv preprint arXiv:2001.10631},
  year={2020}
}

@incollection{tropp2015convex,
  title={Convex recovery of a structured signal from independent random linear measurements},
  author={Tropp, Joel A},
  booktitle={Sampling Theory, a Renaissance},
  pages={67--101},
  year={2015},
  publisher={Springer}
}

@incollection{gordon1988milman,
  title={On Milman's inequality and random subspaces which escape through a mesh in $\mathbb{R}^n$},
  author={Gordon, Yehoram},
  booktitle={Geometric aspects of functional analysis},
  pages={84--106},
  year={1988},
  publisher={Springer}
}

@article{ding2020average,
  title={The average-case time complexity of certifying the restricted isometry property},
  author={Ding, Yunzi and Kunisky, Dmitriy and Wein, Alexander S and Bandeira, Afonso S},
  journal={arXiv preprint arXiv:2005.11270},
  year={2020}
}

@article{baraniuk2008simple,
  title={A simple proof of the restricted isometry property for random matrices},
  author={Baraniuk, Richard and Davenport, Mark and DeVore, Ronald and Wakin, Michael},
  journal={Constructive Approximation},
  volume={28},
  number={3},
  pages={253--263},
  year={2008},
  publisher={Springer}
}

@article{candes2008restricted,
  title={The restricted isometry property and its implications for compressed sensing},
  author={Candes, Emmanuel J},
  journal={Comptes rendus mathematique},
  volume={346},
  number={9-10},
  pages={589--592},
  year={2008},
  publisher={Elsevier}
}

@article{jalal2020robust,
  title={Robust compressed sensing using generative models},
  author={Jalal, Ajil and Liu, Liu and Dimakis, Alexandros G and Caramanis, Constantine},
  journal={Advances in Neural Information Processing Systems},
  year={2020}
}

@book{foucart2013mathematical,
  title={A Mathematical Introduction to Compressive Sensing},
  author={Foucart, Simon and Rauhut, Holger},
  year={2013},
  publisher={Springer Science \& Business Media}
}

@article{van2018compressed,
  title={Compressed sensing with deep image prior and learned regularization},
  author={Van Veen, Dave and Jalal, Ajil and Soltanolkotabi, Mahdi and Price, Eric and Vishwanath, Sriram and Dimakis, Alexandros G},
  journal={arXiv preprint arXiv:1806.06438},
  year={2018}
}

\newpage
\appendix
\section{Appendix}

\begin{proof}[Proof of Theorem \ref{main}]
Since $T$ is closed, there exists a (possibly not unique) point $\mathbf{x}_0 = \mathrm{argmin}_{\mathbf{x}' \in T} ||\mathbf{x} - \mathbf{x}'||_2$. Let $\mathbf{r} \coloneqq \mathbf{x} - \mathbf{x}_0$ and $\mathbf{h} \coloneqq \xhat - \mathbf{x}_0 \in T-T$. We have

    \begin{align}
        ||\mathbf{B} \mathbf{A}\mathbf{h} - (\mathbf{B}\mathbf{A}\mathbf{r}+\mathbf{w})||_2^2 & = ||\mathbf{B}\mathbf{A}\xhat - \mathbf{B}\mathbf{A}\mathbf{x}_0 - \mathbf{B}\mathbf{A} \mathbf{r} - \mathbf{w}||_2^2 \\
        &= ||\mathbf{B} \mathbf{A} \xhat - (\mathbf{B} \mathbf{A} \mathbf{x} + \mathbf{w})||_2^2 \\
        &= ||\mathbf{y}-\mathbf{B} \mathbf{A}\xhat||_2^2 \\
        & \leq ||\mathbf{y}-\mathbf{B}\mathbf{A}\mathbf{x}_0||_2^2 + \epsilon^2 = ||\mathbf{B} \mathbf{A}\mathbf{r}+\mathbf{w}||_2^2 + \epsilon^2.
    \end{align}

Expanding the LHS and rearranging yields
\begin{equation} \label{one}
        ||\mathbf{B} \mathbf{A} \mathbf{h}||_2^2  \leq 2\langle \mathbf{B} \mathbf{A} \mathbf{h}, \mathbf{w} \rangle + 2\langle \mathbf{B} \mathbf{A} \mathbf{h}, \mathbf{B} \mathbf{A} \mathbf{r} \rangle + \epsilon^2.
\end{equation}
The LHS is concentrated about $||\mathbf{B}||_F^2 \cdot ||\mathbf{h}||_2^2$. In fact, by Theorem 1.1 of \cite{jeong2020subgaussian} we can write
\begin{equation}
    ||\mathbf{B} \mathbf{A}\mathbf{h}||_2^2 \geq ||\mathbf{h}||_2^2 \Big[ ||\mathbf{B}||_F - CK\sqrt{\log K} ||\mathbf{B}|| \big(w(T')+ \alpha \cdot \mathrm{rad}(T')\big) \Big]^2
\end{equation}
with probability at least $1-3e^{-\alpha^2}$. Choosing $\alpha = w(T')$ and using the fact that $\mathrm{sr}(\mathbf{B}) \gg K^2 \log K \cdot w^2(T')$, we get
\begin{equation} \label{lb}
    ||\mathbf{B} \mathbf{A}\mathbf{h}||_2^2 \gtrsim ||\mathbf{B}||_F^2 \cdot ||\mathbf{h}||_2^2,
\end{equation}
with probability at least $1-3e^{-w^2(T')}$.

To bound the first term on the RHS, define the random process $X_{\mathbf{t}} \coloneqq \langle \mathbf{B} \mathbf{A} \mathbf{t}, \mathbf{w} \rangle$, for $\mathbf{t} \in T'$. Recall that $||\cdot||_{\psi_2} = \sup_{||\mathbf{u}||_2 = 1} ||\langle \mathbf{u}, \cdot \rangle||_{\psi_2}$. So,  $||X_{\mathbf{t}}-X_{\mathbf{s}}||_{\psi_2} = ||\langle \mathbf{t} - \mathbf{s}, \mathbf{A}^\top \mathbf{B}^\top \mathbf{w} \rangle||_{\psi_2} \leq ||\mathbf{t} - \mathbf{s}||_2 ||\mathbf{A}^\top \mathbf{B}^\top \mathbf{w}||_{\psi_2} \lesssim K ||\mathbf{B}^\top \mathbf{w}||_2 ||\mathbf{t} - \mathbf{s}||_2 \leq K ||\mathbf{B}|| \cdot ||\mathbf{w}||_2 \cdot ||\mathbf{t} - \mathbf{s}||_2$. Therefore, by Talagrand's comparison inequality (Exercise 8.6.5 of \cite{vershynin}),
\begin{equation}
    \sup_{\mathbf{t} \in T'} |X_{\mathbf{t}}| = \sup_{\mathbf{t} \in T'} \langle \mathbf{B} \mathbf{A} \mathbf{t}, \mathbf{w} \rangle \lesssim K ||\mathbf{B}|| \cdot ||\mathbf{w}||_2 \big( w(T') + \beta \cdot \mathrm{rad}(T') \big)
\end{equation}
with probability at least $1-2e^{-\beta^2}$. Again, by setting $\beta = w(T')$ we get
\begin{equation} \label{ub}
    \langle \mathbf{B} \mathbf{A}\mathbf{h}, \mathbf{w} \rangle \leq ||\mathbf{h}||_2 \cdot \sup_{\mathbf{t} \in T'} \langle \mathbf{B} \mathbf{A} \mathbf{t}, \mathbf{w} \rangle \lesssim ||\mathbf{h}||_2 \cdot K||\mathbf{B}|| \cdot ||\mathbf{w}||_2 w(T')
\end{equation}
with probability at least $1-2e^{-w^2(T')}$.

Now let us bound the second term on the RHS. By Cauchy-Schwarz inequality, $\langle \mathbf{B} \mathbf{A} \mathbf{h}, \mathbf{B} \mathbf{A} \mathbf{r} \rangle \leq ||\mathbf{B} \mathbf{A} \mathbf{h}||_2 ||\mathbf{B} \mathbf{A} \mathbf{r}||_2$. Theorem 1.1 of \cite{jeong2020subgaussian} implies that $||\mathbf{B} \mathbf{A}\mathbf{h}||_2 \lesssim ||\mathbf{B}||_F ||\mathbf{h}||_2$ with probability at least $1-3e^{-w^2(T')}$, considering that $\mathrm{sr}(\mathbf{B}) \gg K^2 \log K \cdot w^2(T')$. The same theorem can be used on singleton $\{\mathbf{r}\}$ to bound $||\mathbf{B}\mathbf{A}\mathbf{r}||_2$. Since $w(\{\mathbf{r}\}) = 0$ and $\mathrm{rad}(\{\mathbf{r}\}) = ||\mathbf{r}||$, we get

    \begin{align}
        ||\mathbf{B}\mathbf{A}\mathbf{r}||_2 & \leq ||\mathbf{B}||_F ||\mathbf{r}||_2 + CuK\sqrt{log K} ||\mathbf{B}|| ||\mathbf{r}||_2 \\
        & = ||\mathbf{B}||_F ||\mathbf{r}||_2 \Big(1+ \frac{CuK\sqrt{log K}}{\sqrt{\mathrm{sr}(\mathbf{B})}} \Big),
    \end{align}

with probability at least $1-3e^{-u^2}$. Choosing $u = w(T')$ and using $\mathrm{sr}(\mathbf{B}) \gg K^2 \log K \cdot w^2(T')$ yields $||\mathbf{B}\mathbf{A}\mathbf{r}||_2  \lesssim ||\mathbf{B}||_F ||\mathbf{r}||_2$, with probability at least $1-3e^{-w^2(T')}$.
Thus,
\begin{equation} \label{cross}
    \langle \mathbf{B} \mathbf{A} \mathbf{h}, \mathbf{B} \mathbf{A} \mathbf{r} \rangle \lesssim  ||\mathbf{B}||_F^2 \cdot ||\mathbf{h}||_2 ||\mathbf{r}||_2,
\end{equation}
with probability at least $1-3e^{-w^2(T')}-3e^{-w^2(T')} \geq 1-6e^{-w^2(T')}$.

Combining equations (\ref{one})-(\ref{cross}) gives us
\begin{equation}
    ||\mathbf{B}||_F^2 ||\mathbf{h}||_2^2 - C ||\mathbf{h}||_2  \big( K||\mathbf{B}|| w(T') ||\mathbf{w}||_2 +  ||\mathbf{B}||_F^2 ||\mathbf{r}||_2 \big) - \epsilon^2 \leq 0,
\end{equation}
which implies

    \begin{align}
        ||\mathbf{h}||_2 & \leq \frac{C(K||\mathbf{B}|| w(T') ||\mathbf{w}||_2 +  ||\mathbf{B}||_F^2 ||\mathbf{r}||_2)}{2||\mathbf{B}||_F^2} \\
        &+ \sqrt{\Big( \frac{C(K||\mathbf{B}|| w(T') ||\mathbf{w}||_2 +  ||\mathbf{B}||_F^2 ||\mathbf{r}||_2)}{2||\mathbf{B}||_F^2} \Big)^2 + \frac{\epsilon^2}{||\mathbf{B}||_F}} \\
        & \leq \frac{C(K||\mathbf{B}|| w(T') ||\mathbf{w}||_2 +  ||\mathbf{B}||_F^2 ||\mathbf{r}||_2)}{||\mathbf{B}||_F^2} + \frac{\epsilon}{||\mathbf{B}||_F} \\
        & \lesssim \frac{K w(T')}{||\mathbf{B}||_F \sqrt{\mathrm{sr}(\mathbf{B})}} ||\mathbf{w}||_2 + \frac{\epsilon}{||\mathbf{B}||_F} + ||\mathbf{r}||_2,
    \end{align}

with probability at least $1-3e^{-w^2(T')}-2e^{-w^2(T')} - 6e^{-w^2(T')} = 1-11e^{-w^2(T')}$. Finally,

    \begin{align}
        ||\xhat - \mathbf{x}||_2 & \leq ||\xhat - \mathbf{x}_0||_2 + ||\mathbf{x} - \mathbf{x}_0||_2  = ||\mathbf{h}||_2 + ||\mathbf{r}||_2 \\
        & \lesssim \frac{K w(T')}{||\mathbf{B}||_F \sqrt{\mathrm{sr}(\mathbf{B})}} ||\mathbf{w}||_2 + \frac{\epsilon}{||\mathbf{B}||_F} + ||\mathbf{r}||_2,
    \end{align}

with the aforesaid probability.
\end{proof}

\begin{lemma}\label{quadrant}
A $k$-dimensional subspace in $\mathbb{R}^n$ intersects at most $2^k {n \choose k}$ different orthants.
\end{lemma}

\begin{lemma}\label{min}
Let $D \subset \mathbb{R}^n$ be a $k$-dimensional subspace, and $Q$ be an orthant with $q$ number of positive (and $n-q$ negative) coordinates. Then $\sigma(D \cap Q)$ is contained in a subspace of dimension at most $\min(k,q)$.
\end{lemma}

\begin{proof}[Proof of Lemma \ref{count}]
By the rank-nullity theorem, the subspace $\mathrm{im}(\mathbf{A}_1) \subset \mathbb{R}^{p_1}$ is at most $k$-dimensional. By Lemma \ref{quadrant}, this particular subspace hits at most $2^k {p_1 \choose k}$ different orthants. Therefore, by Lemma \ref{min}, $\sigma(\mathrm{im}(\mathbf{A}_1))$ is contained in a union of at most $2^k {p_1 \choose k}$ subspaces of dimension at most $k$. Each of these subspaces, when multiplied by $\mathbf{A}_2$, is mapped to another subspace of dimension at most $k$ in $\mathbb{R}^{p_2}$. Thus, the linear transformations do not increase the number nor the dimension of the subspaces. Hence, at each layer $i$, every subspace breaks into $2^k {p_i \choose k}$ subspaces, at most. All in all, after $d$ layers, we end up having a union of $N$ subspaces of dimension at most $k$, where
\begin{equation}
N \leq \prod_{i=1}^{d} 2^k {p_i \choose k} \leq \prod_{i=1}^{d} \Big(\frac{2ep_i}{k}\Big)^k = \Big[\big(\frac{2e}{k}\big)^{d}\Big(\prod_{i=1}^{d}p_i\Big)\Big]^k.
\end{equation}
Thus,
\begin{equation}
    T = \mathrm{ran}(G) \subset \bigcup_{i=1}^{N} E_i,
\end{equation}
and
\begin{equation}
    T-T \subset \bigcup_{i=1}^{N} \bigcup_{j=1}^{N} (E_i - E_j) = \bigcup_{1 \leq i<j \leq N} (E_i + E_j),
\end{equation}
where $\dim(E_i) \leq k$ and $\dim(E_i + E_j) \leq 2k$.
\end{proof}

\begin{lemma} \label{concentration}
Let $\mathbf{g} \sim \mathcal{N}(0, \mathbf{I}_n)$ and $f: \mathbb{R}^n \rightarrow \mathbb{R}$ be a Lipschitz function. Then
\begin{equation}
    ||f(\mathbf{g}) - \mathbb{E} f(\mathbf{g})||_{\psi_2} \lesssim ||f||_{\mathrm{Lip}}.
\end{equation}
\end{lemma}

\begin{lemma} \label{max-subg}
Let $X_1, X_2, \cdots, X_N$ be sub-gaussian random variables with $K = \max_{i} ||X_i||_{\psi_2}$. Then
\begin{equation}
    \mathbb{E} \max_{i} |X_i| \lesssim K \sqrt{\log N}.
\end{equation}
\end{lemma}

\begin{lemma} \label{union-mean-width}
Let $T_1, \cdots, T_N \subset \mathbb{S}^{n-1} \subset \mathbb{R}^n$. Then
\begin{equation}
    w \big( \bigcup_{i=1}^{N} T_i \big) \lesssim \max_{i} w(T_i) + \sqrt{\log N}.
\end{equation}
\end{lemma}

\begin{proof}[Proof of Lemma \ref{union-mean-width}]
Let $\mathbf{g}$ be a standard Gaussian vector and define $z_i \coloneqq \sup_{\mathbf{x} \in T_i} \langle \mathbf{x}, \mathbf{g} \rangle - w(T_i)$. Equivalently, let $f_i(\cdot) \coloneqq \sup_{\mathbf{x} \in T_i} \langle \mathbf{x}, \cdot \rangle$ and check that $z_i = f_i(\mathbf{g}) - \mathbb{E}f_i(\mathbf{g})$. Lemma \ref{concentration} implies $||z_i||_{\psi_2} = ||f_i(\mathbf{g}) - \mathbb{E}f_i(\mathbf{g})||_{\psi_2} \lesssim ||f_i||_{\mathrm{Lip}} = 1$. Then, by Lemma \ref{max-subg}, we have $\mathbb{E} \max_{i} z_i \lesssim \sqrt{\log N} \max_{i} ||z_i||_{\psi_2} \lesssim \sqrt{\log N}$. Hence,

\begin{align}
    w\Big(\bigcup_{i=1}^{N} T_i \Big) & = \mathbb{E} \sup_{\mathbf{x} \in \cup T_i} \langle \mathbf{x}, \mathbf{g} \rangle \\
    & = \mathbb{E} \max_{i} \sup_{\mathbf{x} \in T_i} \langle \mathbf{x}, \mathbf{g} \rangle \\
    & = \mathbb{E} \max_{i} \big(z_i + w(T_i)\big) \\
    & = \max_{i} w(T_i) + \mathbb{E} \max_{i} z_i \\
    & \lesssim \max_{i} w(T_i) + \sqrt{\log N}.
\end{align}

\end{proof}

\begin{proof} [Proof of Proposition \ref{width}]

By Lemma \ref{count}, $T-T \subset \bigcup_{i=1}^{{N \choose 2}} F_i$, where $\mathrm{dim}(F_i) \leq 2k$. Thus,
\begin{equation}
    w\big((T-T) \cap \mathbb{S}^{n-1}\big) \leq w\big( \bigcup_{i=1}^{N \choose 2} (F_i \cap \mathbb{S}^{n-1}) \big).
\end{equation}
Using Lemma \ref{union-mean-width} and Lemma \ref{count} respectively, then we have

    \begin{align}
        w\big( \bigcup_{i=1}^{N \choose 2} (F_i \cap \mathbb{S}^{n-1}) \big) & \lesssim \max_{i} w(F_i \cap \mathbb{S}^{n-1}) + \sqrt{\log {N \choose 2}} \\
        & \leq \max_{i} \sqrt{\mathrm{dim}(F_i)} + \sqrt{2 \log N} \\
        & \leq \sqrt{2k} + \sqrt{2kd \log(\frac{2ep'}{k})} \\
        & \lesssim \sqrt{kd \log (\frac{p'}{k})}.
    \end{align}

\end{proof}

\end{document}